# PHILOSOPHICAL SMOKE SIGNALS: THEORY AND PRACTICE IN INFORMATION SYSTEMS DESIGN

David King and Chris Kimble

David King, University of Sheffield, Department of Computer Science, 211 Portobello Street, Sheffield, S1 4DP.  david@dcs.shef.ac.uk  +44 1142 221 831

Chris Kimble, University of York, Department of Computer Science, Heslington, York, YO10 5DD.  kimble@cs.york.ac.uk  +44 1904 433 380

**Abstract**
Although the gulf between the theory and practice in Information Systems is much lamented, few researchers have offered a way forward except through a number of (failed) attempts to develop a single systematic theory for Information Systems.  In this paper, we encourage researchers to re-examine the practical consequences of their theoretical arguments.  By examining these arguments we may be able to form a number of more rigorous theories of Information Systems, allowing us to draw theory and practice together without undertaking yet another attempt at the holy grail of a single unified systematic theory of Information Systems.

## 1  Background
**Theory and Practice**
Although the foundations of Information Systems have been much debated over the last 40 years, the practical value of these debates has often been questioned (Bacon and Fitzgerald, 2001; Oettinger, 1964).  This has lead to something of a gulf between Information Systems theory and practice (Farhoomand and Drury, 1999; Glass, 1996).  However, if we are to address either theoretical or practical problems, we must also deal with the relationship between the two.

For instance, recent papers from the 'theory' side of Information Systems research have pushed for the formation of a 'scientific' foundation for Information Systems (Farhoomand, 1987; Khazanchi and Munkvold, 2000).  In one sense, pushing for a 'scientific' foundation for Information Systems is not a bad idea.  The problems start when we ask, "what kind of science might underpin Information Systems"?

In recent years, we seem to have seen a concerted push to move Information Systems to the same 'scientific foundation' as Computer Science.  The wisdom of such a move is questionable.  We could for example point out the scientific nature of Computer Science has been a source of active debate for over 50 years (Dijkstra, 2001; McGuffee, 2000; Proulx et al, 1996).  However, a more serious criticism would be that the original impetus for creating a separate field of Information Systems was the sense the context of Computer Science and Information Systems were clearly distinct (Davis et al, 1996.  For instance, in the Association for Computational Machinery (Jay

et al, 1982, p. 784) argued when drawing up a curriculum for undergraduate studies, that:

*"The IS curriculum teaches information system concepts and processes within two contexts, organization functions and management knowledge, and technical information systems knowledge"*

by contrast:

*"… Computer Sciences [sic] tends to be taught within an environment of mathematics, algorithms and engineering technology."*

**The Context of Information Systems**
When researchers form theories in Information Systems, they must be aware of the practical context of their theories. Hence, the properties of "*organization functions*" or "*technical information systems knowledge*" are not simply a matter of academic debate. Rather, these theoretical assumptions have deep implications for anyone using these theories in practice.

In this paper, we will explore one facet of the practical implications of such theoretical assumptions by examining two problems in software design. One problem concerns the nature of theories about knowledge and one the nature of theories about the world. Since many Information Systems require (or at least use) a software component, we hope this focus will not overshadow the broader arguments of the paper.

Software design presents many interesting opportunities to study the practical consequences of theoretical assumptions (King and Kimble, 2004a). Software designers' work in the context assumed by researchers in Information Systems: the context of "*organization functions and management knowledge and technical information systems knowledge*". In contrast, computer programs have a distinct mathematical and philosophical heritage: one that has remained fixed since at least the 1930s. Consequently, program designers must work in the context assumed by researchers in Computer Science: the context of "mathematics, algorithms and engineering technology".

In short, program designers must use closed, consistent and complete characterisations of the world if the final design is to be of any use. Software designers, by contrast, can (and usually do) deal with open, inconsistent and incomplete characterisations of the world. Exactly how the two worlds of software and program design can be reconciled is a theoretical question: but the answers are of vital importance to any practical solution.

**Terminology**
In this paper, we will confine our examination of software design to just two debates: one over the term information and another over the term knowledge. Our aim in this paper is not to establish incontrovertible definitions for either of these two terms; rather we are interested in the arguments used by others in trying to establish such definitions. Only by understanding these arguments can we gain an insight into the practical consequences of theoretical assumptions.

However, we do have to start somewhere. In this paper, we frame our discussion by reference to three important concepts: descriptions, representations and reality. Recognising the impossibility of 'philosophy-free' definitions, we define three concepts (somewhat arbitrarily) by reference to the Oxford English Dictionary. We hope by doing this our readers will forgive the emphasis given to particular schools of philosophical thought during the definition of these three concepts.

For the first of our three concepts, we begin with the unperceived universe: a universe beyond the reach of any individual designer. This unperceived universe we will call **reality**, defined as "... *that which underlies and is the truth of appearances or phenomena*" (Simpson and Weiner, 1989). In common with academic philosophical discussions, our emphasis in the concept of reality is centred on what really exists. In particular, our use of the term 'reality' does not refer to the existence perceived by any individual designer (Audi, 1995). Nonetheless, the perceptions of an individual designer can be important to the success of the software design (Sommerville and Sawyer, 1997). Often, it is only by capturing these viewpoints that a designer can fully understand the needs of all the users (Graham, 1996). Thus in software design, in addition to 'unperceived reality', we need a notion of 'perceived reality'; we will call this perceived reality the **representation**. The representation can be thought of as the mental model used by an individual designer, or more simply, the world imagined by the designer. The representation can therefore be defined as "*[t]he operation of the mind in forming a clear image or concept*" (Simpson and Weiner, 1989). However, these mental models are not directly transmissible. Instead, some intermediate form is needed; we will call this form the **description**. For now, we will simply define descriptions as "*[a] statement which describes, sets forth, or portrays*" (Simpson and Weiner, 1989).

## 2 Knowledge
**Epistemological Positions**
Having defined our three concepts, descriptions, representations and reality, we will now examine the first of our theoretical debates, the debate over the term 'knowledge'. In philosophical discussions, debates over the term knowledge fall under the realm of epistemological arguments. Within the field of epistemology, researchers in software design have often exploited two contrasting extreme viewpoints. The first is the rationalist position, which argues that knowledge is the product of thought and reason (Jack, 1993b). This forms the foundation for software design methods that see the concepts 'representation' and 'description' as being equivalent. For instance, software design methods using formal theory frequently draw an equivalence between software and program design methods. Early software design methods, such as the Structured System Design Methods used in the 1960s and 1970s, borrow from rationalist arguments (Connell and Shafer, 1989).

At the other extreme to rationalist arguments lies the empiricist position. Researchers borrowing from the empiricist position argue that knowledge is the result of observation and experience (Jack, 1993a). Thus, software designers following the empiricist position do not see the concepts 'representation' and 'description' as equivalent but do see the concepts of reality and the representation as being equivalent. For example, researchers in object-oriented software design methods borrow from empiricist arguments by making the distinction between objects as they exist in reality, and objects perceived by the software designers, equivalent. Software design methods proposed by researchers using the early forms of general systems theory, for instance Peter Checkland's Soft System Methodology, also borrow from empiricist arguments Checkland, 1981).

As philosophical arguments, rationalist and empiricist positions represent two extreme positions. Most philosophical schools take a position somewhere in-between these two. Nonetheless, these philosophical arguments are carefully crafted. Researchers in software design, though, have tended to be less philosophically rigorous. While this may seem a trivial point, when we try to apply theories, such philosophical lapses have a tendency to bite back.

What happens, for instance, if we try to define and apply a formal theory of object-oriented design? Can we simply conflate arguments from both the rationalist and the empiricist schools and create a suitable theoretical foundation for a new design method? Alternatively, do the different undercurrents in the two schools of philosophy create a tension leading to unsolvable problems when we try to define what 'knowledge' means in the context of our new theory?

**Epistemology in Software Design**
Practically, while rationalist positions consider the representation to be stable, empiricist positions consider the representation to the unstable. The stability, or otherwise, of the representation creates a fundamental problem for applied theories of software design borrowing epistemological arguments from different schools of thought. Take, for instance, the object-oriented design of a juice plant control system shown in Figure 1. We will not go into a detailed discussion of the graphical syntax of Figure 1, rather we will only consider the practical problems of a designer trying to form a description of the cooking tank.

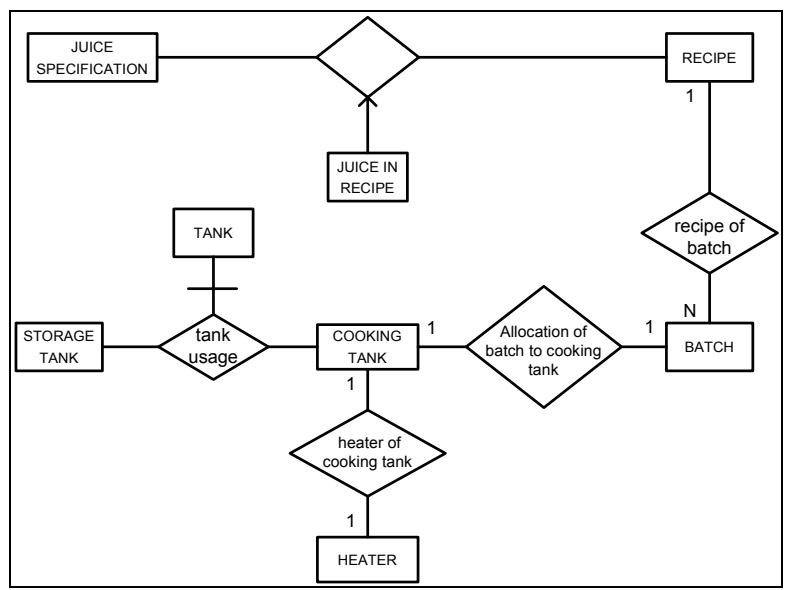

(a) Fragment of an Entity-Relationship Diagram of the data manipulated by a Juice Plant Control System

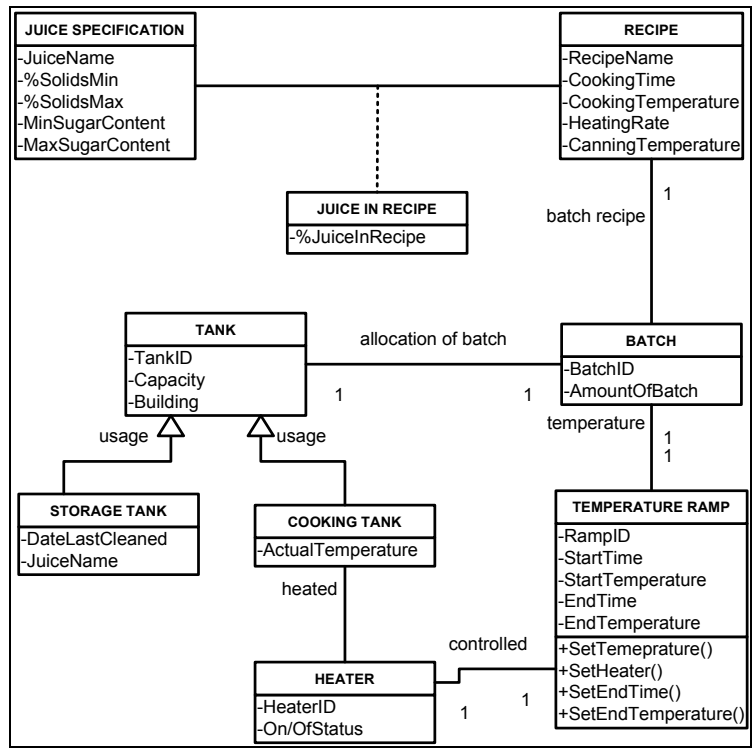

(b) Fragment of a UML Class Diagram of the Juice Plant Control System

Figure 1: Fragments of an Object-Oriented Design for a Juice Plant Control System (Wieringa, 1998 - Figures 5 and 6)

According to empiricist arguments, the designer first forms a representation of the cooking tank from the experience of the cooking tank, as pictured in Figure 2(a). From this experience, the designer can then form a description of the representation, as illustrated in Figure 2(b). However, simply forming the description from representation also counts as experience. Consequently, the original representation must change, to take into account this new experience, as shown in Figure 2(c). In the empiricist view, the representation and the description are always *out of step* and so cannot be equated.

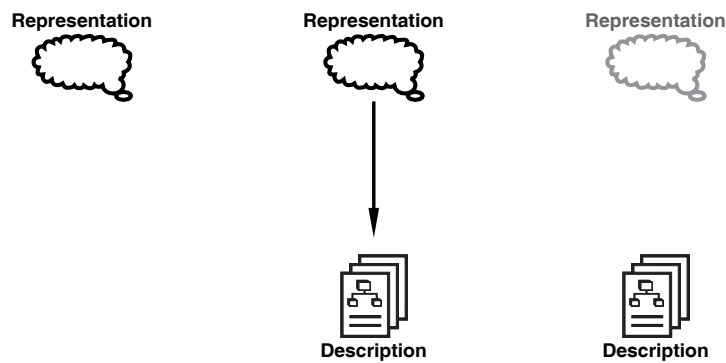

(a) Before a Change    (b) During a Change    (c) After a Change

Figure 2: The Empiricist Relationship between Description and Representation

To get around this instability, the rationalist approach argues that reason is independent of experience. If the principles of reason are independent of experience, then it is possible to bring the description and representation back in step with one another. In the rationalist view, the designer starts with the description (not the representation) as illustrated in Figure 3(a). By applying the principles of reason (which are independent of experience), the designer can then form a representation from the description, illustrated by Figure 3(b). Unlike the empiricist position, there is nothing that now forces a change to either the description or the representation. Thus, the description and representation remain in step, even after a change in the description, as indicated in Figure 3(c).

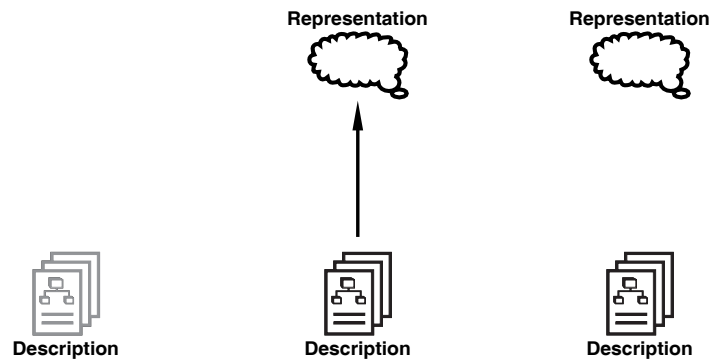

(a) Before a Change   (b) During a Change   (c) After a Change

Figure 3: The Rationalist Relationship between Description and Representation

Once consequence of the rationalist arguments is the assumption that all designers are conscious of their own representations. In other words, nothing exists in any designers' representation that cannot be represented in a description created by rational means.

Knowledge created though the full consciousness of an individual mind is called *explicit knowledge*. By contrast, empiricist approaches also recognise the existence of knowledge that a designer is not consciously aware of. This type of knowledge is called *tacit knowledge* and cannot be turned into a description because the designer has no direct awareness of it. Most empiricist approaches claim (Audi, 1995) that:

"*Much of our knowledge is tacit: it is genuine but we are unaware of the relevant states of knowledge, even if we can achieve awareness upon suitable reflection.*"

Current evidence suggests that adequately dealing with tacit knowledge is a real problem (Desouza, 2003, Shipman and Marshall, 1999). Yet, if we adopt a strict rationalist position, tacit knowledge should not exist, since such knowledge is defined by rational articulation. We can incorporate tacit knowledge into approaches leaning towards rationalist positions (Reeves and Shipman, 1996), but while this makes it easier to form software descriptions, forming the corresponding program descriptions becomes harder.

**Defining Knowledge**
Both the rationalist and empiricist positions offer attractive features as theoretical foundations for methods of software design. However, neither approach offers a definition of knowledge that is easily reconcilable with the other. In the rationalist approach, it is easier to create descriptions that are clear and unambiguous, because all designers must interpret the description in the same way and all designers have full awareness of their own representations. This eliminates any tacit knowledge that might lie behind the design. By eliminating tacit knowledge, the practical problem of creating descriptions of Information Systems should be eased.

However, if we find tacit knowledge is important, moving to empiricist arguments may help. Software design in this view is often more dynamic, with new experiences constantly forcing a change in the descriptions of the software design. However, if we move to empirical theories, we have a new problem of design consistency. Empiricist approaches expect different designers to have different representations of the design - which makes it impossible to demonstrate that the final design is consistent.

In short, adopting theories at either extreme will lead to practical problems with either the use of tacit knowledge or with design consistency. We cannot easily short-circuit our difficulties by conflating theories from both approaches, since these theories rely on incommensurable notions of knowledge. If we find problems with design consistency, using theories based on rationalist arguments may help. Similarly, if we find tacit knowledge is important, using empiricist theories may help. However, neither approach will solve both problems simultaneously: we are always left with an uncomfortable compromise.

## 3  Information
**Ontological Positions**
Leaving aside the problem of knowledge for a moment, we can see the same kinds of practical and theoretical issues played out when dealing with the (apparently) more straightforward problem of information. Given the strong theoretical links between formal theory and information theory, most researchers in software design methods have assumed that the theoretical definition of information has few, if any, practical consequences. Certainly, researchers show far more agreement over the definition of the term 'information', compared to the lack of consensus in the debate over the term 'knowledge' (Meadow and Yuan, 1997).

Nonetheless, software design methods still draw on distinct ontological arguments when trying to define 'information'. In philosophy, the problems of ontology focus on questions on the nature of reality. As with epistemology, there are two broad extremes woven into most philosophical positions. The first forms the foundation for realist arguments, which "… *emphasises that truth is possible: beliefs are testable against 'reality' and that reality is 'knowable'*" (Duro et al, 1993). In other words, realist arguments emphasise that reality is independent of a designer. Software design methods from both the formal and object-oriented strands of research often show examples of the influence of realist ontological arguments.

Nonetheless, realist ontological arguments are not completely ubiquitous in software design. Older structured software design methods for procedural languages often use arguments from the other broad ontological extreme: anti-realism. Anti-realist arguments claim that the perception of reality is so bound to the observing mind that:

"*Even if we could impart the highest degree of clearness to our intuition, we should not come one step nearer to the nature of objects by themselves.*" (Kant, 1966, p 36)

In software design, a common problem lies in trying to reconcile the logical perfection of the program design with the behaviour of the real world. Structured software design methods address this problem by separating the logical design (the representation) from the physical design (modelling reality).

**Ontology in Software Design**
Most modern software design methods prefer seamless design and try not to separate the logical and physical designs (Jackson, 2001). In addition, realist ontological arguments offer many attractions to researchers trying to develop the systematic theories required by seamless software design. By emphasising the independence of reality in the realist position, it is always possible to reconcile a representation to reality. This becomes an equivalence between a designer's representation and reality.

For example, using the cooking tank problem, the perception of the designer (the representation) is fundamentally different to the real object of the cooking tank. This is situation is pictured in Figure 4(a). If a designer now witnesses a change in the properties of a cooking tank, every detail of that change is apparent to the designer. Moreover, not every detail of that change depends on the perception of the designer. This means that the designer can update their representation to bring it back into line with the new reality as shown in Figure 4(b). After both the representation and reality have changed, the independence of the two means that neither need undergo any further changes. Hence, both reality and the representation *stay in step* and an equivalence between the two is preserved as pictured in Figure 4(c).

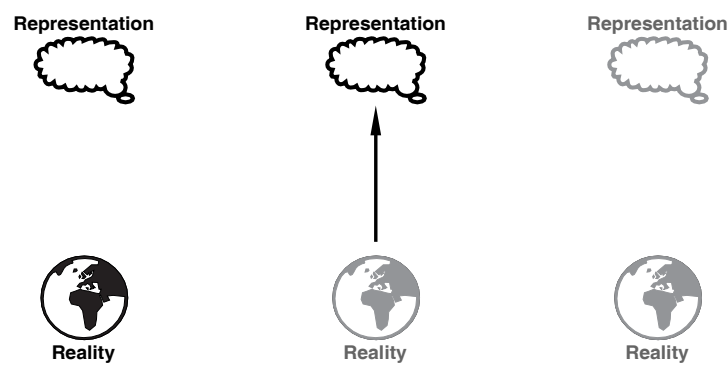

(a) Before a Change   (b) During a Change   (c) After a Change

Figure 4: The Realist Relationship between Reality and Representation

In contrast, according to anti-realist arguments, there is no independence between reality and a designer's representation. In the anti-realist arguments, 'reality' cannot be said to exist unless it is perceived. Consequently, we cannot maintain an equivalence between these two concepts.

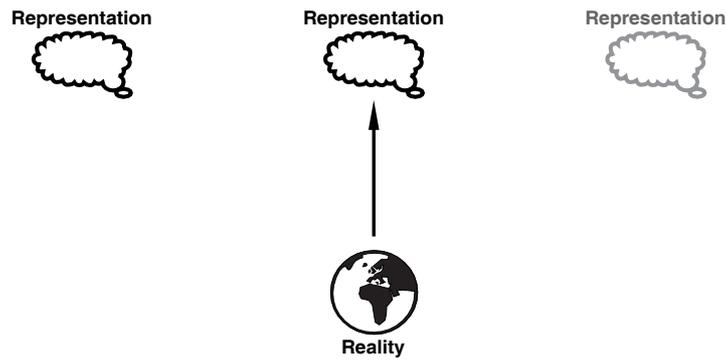

(a) Before a Change   (b) During a Change   (c) After a Change

Figure 5: The Non-Realist Relationship between Reality and Representation

When dealing with the cooking tank problem, only the representation held by an individual designer is relevant. Whatever 'reality' may or may not be is immaterial as it has no influence on the designer, as shown in Figure 5(a). Only when a designer changes their perception of reality, for instance in response to a change in the character of a cooking tank, is the representation changed, as shown in Figure 5(b). After this change, 'reality' ceases to have an influence on the representation of the designer, as pictured in Figure 5(c). This means that there can be no equivalence between representation and reality. It also means that the individual representations of designer are unique to those designers. There is no common point of reference as there is in the realist position. Again, this means that demonstrating a consistency between a representation and reality becomes impossible in the anti-realist position.

**Defining Information**
In recent years, most new software design methods prefer a realist ontological position as realism offers three useful claims (Audi, 1995):

1. There are real chunks of the world.

2. These chunks exist independently of any designers experience and knowledge of them.

3. The chunks have properties and enter into relations independently of the concepts used by designers to understand them, or of the language designers use to describe them.

Together, these three properties allow the assumption that reality forms an independent point of reference. With an independent frame of reference, designers can increase the overall consistency of the design, even if the description is unstable. In particular, realism offers a way of defining information *directly* from properties of reality. If 'data' is formed from reality, we can then see 'information' as simply the transformation of 'data' into the perception of an individual designer.

Nevertheless, this also implies 'data' is somehow an objective reference point for all designers. Older methods of software design often dispute this, recognising a distinction between the 'ideal' flow of data and data transformation as it occurs in the real world. Thus, older methods of software design use anti-realist arguments to separate the description from reality. This means that a clear, unambiguous description is not hampered by the need to demonstrate the same clarity and unambiguity in reality. For structured design methods (and also for methods such as Checkland's Soft Systems Methodology), anti-realism offers another way of emphasising the influence of designers and subjective interpretations of the world. Information in this view is more closely tied with the individual perceptions of designers.

As with epistemological arguments, these different ontological arguments are not easily reconciled with each other. Like epistemological arguments, both anti-realist and realist arguments also have different expectations of an applied theory. Under realist ontological arguments, the structure of theory should be a close analogue of reality. While under anti-realist ontological arguments, mapping theory to reality is non-trivial. Both forms of philosophical argument therefore emphasise the distinctness of design methods following research in different strands.

# 4 Conclusion
**Problems for a Systematic Applied Theory**
When it comes to applying information systems theories, most researchers start by trying to find a single systematic Theory of Information Systems. However, ontological and epistemological differences between the various competing theories pose a considerable challenge.

For instance, in developing a systematic formal theory for object-oriented design, we must somehow deal with the differing epistemological assumptions of formal and object-oriented theory. Commonly, designers using formal software design methods borrow from rationalist epistemological arguments, for instance, by emphasising the role of reason and logic in forming descriptions of reality. By contrast, software designers using object-oriented often emphasise the role of observation in trying to form descriptions of reality. So why can we not simply conflate the two positions and emphasise both reason and observation? The short answer is we could, but we must be very careful in doing so.

Take, for instance, a definition of knowledge in our new formal theory of object-oriented design. Under rationalist epistemological arguments, designers gain knowledge through logical deduction from objective first principles. Further, the representation contains nothing that cannot be expressed equally well in the description. Moreover, the description offers a shared, objective touchstone for all designers. If the distinction between the description and the representation is effectively non-existent, why should our concept of knowledge in this new framework focus on the representation, rather than the description?

In contrast, many object-oriented design methods assume the relationship between the description and representation is real and non-trivial. Borrowing from empiricist epistemological arguments, object-oriented software design methods assume the software description can, and will, get out of step with a designers representation. Indeed whole areas of research into object-oriented design have been devoted to trying to reconcile designer's representation with the descriptions (Berry, 2002; Pancake, 1995; Schwaber, 2002). Hence, for object-oriented software design methods, we may find compelling reasons for using a definition of knowledge that ignores the description and focuses instead on the representation.

**Systematic or Rigorous?**
Returning to the context of Information Systems for a moment, even if the context of Information Systems *is* distinct from the context of Computer Science, we might not need a single a systematic theory. Do we, for instance, require all organisations to be underpinned by a single systematic theory to be able to reason about their behaviour? Such a movement away from systematic theories necessarily invalidate the quest for a scientific underpinning of Information Systems. Economic theories, for example, are widely used to reason about the behaviour of organisations although there is not a single systematic theory of economics.

In software design, we lack good, rigorous, theories of why particular software design methods work or do not work. Developing a deeper understanding of Information Systems (and thereby closing the gulf between theory and practice) may mean a move away from universal systematic theories and towards a range of differing but more rigorous ones - at least in the short term. These rigorous theories might one day join up into a systematic theory of software design, or they may not. Indeed, for software design at least, we have good theoretical reasons for believing a truly systematic theory may never exist (King and Kimble, 2004b).

Developing a range of more rigorous theories may mean abandoning old habits. In software design, researchers often seem to assume that the only valid context for their theories is the same as that adopted by program design. However, formal theories of program design cannot incorporate ambiguity. Although few researchers have looked at the benefits of ambiguous design, some research has highlighted the potential importance of ambiguity. For instance, when Mark Gross and Ellen Do investigated the aids used by designers they commented:

"*We conjecture (based on observation and interviews) that designers prefer to use paper and pencil because it supports ambiguity, imprecision and incremental formalization of ideas as well as rapid exploration of alternatives.*" (Gross and Do, 1996, p 183)

The search for a systematic theory may be a goal for information systems research: but it must not become the only focus of research. Good, rigorous and even scientific theories for Information Systems can be developed without first developing a full blown systematic theory. Similarly, research into software design has often been hampered by the search for the philosopher's stone of a systematic theory of software design (Berry, 2002; Jackson, 2001).

Information system research has a rich heritage of theories drawn from many different disciplines and has stimulated many interesting debates between different areas. Some may feel these debates have proved futile and others argue that they have not yet yielded the promised fruits. Nonetheless, we must not abandon the need for rigorous theories of information systems in favour of quest for an all encompassing systematic one. To do so will only prevent the convergence of theory and practice: a goal we all hope to achieve.